\documentclass[12pt,preprint,showpacs,preprintnumbers,amsmath,amssymb]{revtex4}

\usepackage{graphicx}
\usepackage{dcolumn}
\usepackage{bm}
\usepackage{setspace}

\newcommand{\ttiny}{\fontsize{5.5pt}{6pt}\selectfont}

\begin{document}

\preprint{  arXiv:1609.02512 [hep-ph]  }

\title{ Spectroscopy of singly, doubly, and triply bottom baryons }    

\author{
{    \small    }    
        Ke-Wei Wei$^{1,4,}$\footnote{e-mail: weikw@hotmail.com},
        Bing Chen$^{1,}$\footnote{e-mail: chenbing@shu.edu.cn},
        Na Liu$^{1,3}$, 
        Qian-Qian Wang$^{1}$, %
        and
        Xin-Heng Guo$^{2,}$\footnote{Corresponding author, e-mail: xhguo@bnu.edu.cn}\\
\footnotesize{  $^1$ College of Physics and Electrical Engineering, Anyang Normal University, Anyang 455000, China \\
                $^2$ College of Nuclear Science and Technology, Beijing Normal University, Beijing 100875, China \\
                $^3$ School of Physics and Technology, Nanjing Normal University, Nanjing 210023, China\\
                $^4$ Theoretical Physics Center for Science Facilities, CAS, Beijing 100049, China \\    
             }
}
\affiliation{ }

\date{\today\\}

\begin{abstract}
{       
    Recently, some  singly bottom baryons have been established experimentally, but  none of  doubly or triply bottom baryons has been observed.  
        Under the  Regge phenomenology,  the mass of a ground state unobserved doubly  
    or triply bottom baryon is expressed  as a function of masses of the well established light baryons and singly bottom baryons.
%
%
%
%
%
     Then, the values of Regge slopes and Regge intercepts for baryons containing one, two, or three bottom quarks are calculated.
%
%
    After that, masses of the orbitally excited   singly, doubly, and triply bottom baryons are estimated.    
%
%
%
%
%
%
%
    Our  predictions may be useful for the discovery of these baryons and the $J^P$ assignment of them. 


%
}
\end{abstract}

%

\maketitle

    \begin{spacing}{1.4}
\section{Introduction}

     According to the Particle Data Group's  latest ``Review of Particle Physics'' (RPP) \cite{PDG2015}, 
   some singly bottom baryons ($\Lambda_b$, $\Lambda_{b}(5912)$, $\Lambda_{b}(5920)$,  $\Sigma_b^{}$, $\Sigma_b^{*}$,   $\Xi_b^{}$, $\Xi_b'$, $\Xi_b^{*}$, and $\Omega_{b}^{}$)
   have been  established.  
%
%
%
    However,  none of doubly or triply bottom baryons has been observed until now.
    Therefore,  in the present work,  we will  focus on searching mass relations
  which can be used to express the mass of a doubly or triply bottom baryon as a function of masses of the well established baryons.   
    This is the main motivation of this work.

    It is noted that the Gell-Mann-Okubo formula \cite{Gell-Mann-Okubo} cannot be directly applied  to the charmed and bottom hadrons due to higher-order breaking effects.
  %
    The Regge trajectory ansatz is an effective phenomenological model to study  
 mass relations \cite{Regge2008,  Burakovsky-meson-relation, Burakovsky1997, Kaidalov1982, Wei:2015gsa} and
 mass spectra \cite{De-minLi2004,  Wei:2010zza, Ailin Zhang, non-linear, anisovich, Ebert:2011kk, Ebert:2009ub, Masjuan:2012gc} for mesons and baryons.
    In a previous work \cite{Wei:2015gsa}, a way was proposed to express the mass of a doubly charmed baryons as a function of masses of the well established light baryons and singly charmed baryons.
    In the present work,  under Regge phenomenology, we will  
  express the  mass of an unobserved ground state (the orbital quantum number $L=0$) baryon containing one, two or three bottom quarks   as a function of masses of the well established light baryons and singly bottom baryons.
    The mass values will be given and  compared with those obtained in many other approaches   
  \cite{ Ebert:2011kk, Burakovsky1997, Ponce-1979,   Martynenko:2007je,   Roberts:2007ni, Bjorken:1985ei, Roncaglia:1995az, Ebert:2005xj, He:2004px, Kiselev-etal-2000-2002,
 Kiselev:2000jb, Jia-2006, Hasenfratz:1980ka,       Valcarce:2008dr, Vijande:2014uma, Vijande:2015faa,
    Brown:2014ena, Brown:2014pra, Karliner:2014gca, Ghalenovi:2014swa, Aliev:2012iv,   Wang:2010vn,   Weng:2010rb,    Zhang:2008pm,
 Bagan:1992za, Patel:2008mv, Bernotas:2008bu, Ghalenovi:2011zz,   Kiselev:2001fw, Tang:2011fv, Giannuzzi:2009gh,      LlanesEstrada:2011kc,
 Yoshida:2015tia, Mao:2015gya, Meinel:2010pw,  SilvestreBrac:1996bg, Lewis:2008fu, Zheng:2010zzc, Albertus:2006ya,  AliKhan:1999yb, Ebert:1996ec, Tong:1999qs, Narodetskii:2001bq, Chen:2014nyo, Dhir:2013nka, Faessler:2006ft, Garcilazo:2007eh, Eakins:2012jk, Karliner:2008sv, Capstick:1986bm}.

    The slopes and intercepts of  Regge trajectories are useful for many spectral and nonspectral purposes \cite{Regge2008, De-minLi2004, Burakovsky-para-relation},
  for example,  in the fragmentation \cite{fragmentation-M} and recombination \cite{recombination-M} models.
    Therefore,  
  Regge slopes and Regge intercepts are fundamental constants of hadronic dynamics, perhaps more important than the masses of  particular states \cite{Basdevant1985}.      
    Thus, the determination of Regge slopes and intercepts of hadrons is of great importance since this affords opportunities for  
  a better understanding of the dynamics of strong interactions in the processes of production of charmed and bottom hadrons at high energies and estimates of their production rates \cite{Burakovsky-para-relation}.
    In our previous work in 2008 \cite{Regge2008},    the numerical values of Regge slopes and Regge intercepts for the $\frac{1}{2}^+$ and $\frac{3}{2}^+$ light and charmed baryon trajectories are extracted.
    In Ref. \cite{De-minLi2004}, Li \emph{et al.} gave the values of Regge slopes and intercepts for all $SU(5)$ mesons (involving the $u,d,s,c,b$ quarks).
    In Ref. \cite{Ebert:2011kk}, Ebert \emph{et al.} gave the values of Regge slopes for singly charmed and singly bottom baryons.
%
    In the present work, we will extract Regge  slopes and Regge intercepts for the singly, doubly, and triply bottom baryons
 and calculate  masses of the orbitally excited  baryons ($L$=1,2,3,4) lying on these Regge trajectories.

    The remainder of this paper is organized as follows.
    In Sec. II, under  Regge phenomenology,  
 masses  of  the unobserved ground state singly, doubly, and triply bottom baryons    $\Omega_{b}^{*}$,  $\Omega_{bb}^{*}$,  $\Omega_{bb}$, $\Xi_{bb}^*$, $\Xi_{bb}$, and $\Omega_{bbb}$
  will be given.
 %
    In Sec. III,  
   we will calculate   Regge slopes and Regge intercepts for the $\frac{1}{2}^+$ and $\frac{3}{2}^+$ singly, doubly, and triply bottom baryon trajectories.
%
    After that, masses of the orbitally excited  baryons lying on these  trajectories will be estimated. 
 %
    In Sec. IV,  a short discussion and conclusion will be given.           

\section{Masses of the  unobserved $\Omega_{b}^*$, $\Omega_{bb}^{(*)}$, $\Xi_{bb}^{(*)}$, and $\Omega_{bbb}$ baryons}   

    In this section,    we will first give a short introduction for Regge phenomenology
  and express the  quadratic  masses  of  the ground state unobserved $\Omega_{b}^*$, $\Omega_{bb}^{(*)}$, $\Xi_{bb}^{(*)}$, and $\Omega_{bbb}$ baryons  
 as functions of the quadratic masses of the well established light baryons and singly bottom baryons.  
     After that, the mass values will be given and compared.   

\subsection{Regge phenomenology}   

    Regge theory  is concerned with almost all aspects of strong interactions, including the particle spectra,
  the high energy behavior of scattering amplitudes, and the forces between particles \cite{regge-book}.
    It is known from Regge theory that all mesons and baryons are associated with Regge trajectories
 (Regge poles which move in the complex angular momentum plane as a function of energy)~\cite{Regge-theory}.
      Hadrons lying on the same Regge trajectory which have the same internal quantum numbers are classified into the same family \cite{Chew-Frautschi,regge-book}. 
    Regge trajectories for hadrons can be parameterized as follows:
\begin{equation}
J=\alpha(M)=a(0)+\alpha^\prime M^2, \label{regge1}
\end{equation}
 where $\alpha^\prime$ and $a(0)$ are respectively the slope and intercept of the Regge trajectory on which the particles lie.
     For a baryon multiplet,   Regge intercepts and Regge slopes for different flavors can be related by the following relations (see Ref.~\cite{Regge2008} and references therein):
\begin{equation}
a_{iik}(0)+a_{jjk}(0)=2a_{ijk}(0) \label{intercept2},
\end{equation}
\begin{equation}
\frac{1}{\alpha_{iik}^\prime}+\frac{1}{\alpha_{jjk}^\prime}=\frac{2}{\alpha_{ijk}^\prime}, \label{slope1b}
\end{equation}
 where $i$, $j$, and $k$ denote arbitrary light or heavy quarks.
    Based on Eqs.  (\ref{intercept2}) and (\ref{slope1b}), one can introduce two parameters $\gamma_{x}$ and  $\lambda_{x}$,
\begin{equation}
  \gamma_{x} = \frac{1}{\alpha_{nnx}^\prime}-\frac{1}{\alpha_{nnn}^\prime},
\end{equation}
\begin{equation}
  \lambda_{x} = a_{nnn}(0)-a_{nnx}(0),
\end{equation}
  where $n$  represents the light unflavored quark $u$ or $d$, $x$ denotes  $i$, $j$, or $k$.  %
    Therefore,
\begin{equation} \label{a0ijk}
a_{ijk}(0) =  a_{nnn}(0) - \lambda_{i} - \lambda_{j} -\lambda_{k} ,
\end{equation}
\begin{equation}
\frac{1}{\alpha_{ijk}^\prime} = \frac{1}{\alpha_{nnn}^\prime} + \gamma_{i} + \gamma_{j} + \gamma_{k}     .
\end{equation}
        From  Eq. (\ref{regge1}), one can have
\begin{equation}
J = a_{nnn}(0) + {\alpha_{nnn}^\prime} M_{nnn}^2 ,
\end{equation}
\begin{equation} \label{JMijk}
J = a_{ijk}(0) + {\alpha_{ijk}^\prime} M_{ijk}^2 .
\end{equation}
    %
     From Eqs. (\ref{a0ijk}) - (\ref{JMijk}) , we can have
\begin{equation}
 M_{ijk}^2=(\alpha_{nnn}^\prime M_{nnn}^2+\lambda_i+\lambda_j+\lambda_k)\left(\frac{1}{\alpha_{nnn}'}+\gamma_{i}+\gamma_{j}+\gamma_{k}\right). \label{MB-transf}
\end{equation}
     As demonstrated in Ref. \cite{Regge2008},  in order to evaluate the high-order effects, we  introduce the parameter $\delta$,  
\begin{equation}   \label{delta}
 \delta_{ij,q} \equiv M_{iiq}^2+M_{jjq}^2-2M_{ijq}^2,
\end{equation}
  where $q$ is an arbitrary light or heavy quark.
    Combining Eqs. (\ref{MB-transf}) and (\ref{delta}), we can prove
\begin{equation}
\delta_{ij,q} = M_{iiq}^2+M_{jjq}^2-2M_{ijq}^2
              = 2(\lambda_i-\lambda_j)(\gamma_i-\gamma_j) \label{baryon-equal} .
\end{equation}
    From Eq. (\ref{baryon-equal}), one can  see that $\delta_{ij,q}$ is independent of the  $q$ quark.

      \subsection{  Mass expressions and masses for the   $\frac{3}{2}^+$  $\Omega_{b}^*$, $\Omega_{bb}^{*}$, $\Xi_{bb}^{*}$, and $\Omega_{bbb}$ baryons}  


  For the $\frac{3}{2}^+$ multiplet,
 noticing that $\delta_{ij,q}^{\frac{3}{2}^+}$ is independent of $q$ in the above relation (\ref{baryon-equal}),
  when $i=n$, $j=s$, $q=s,c,b$, Eqs. (\ref{baryon-equal}) can be expressed as
\begin{equation}
\delta_{ns,q}^{\frac{3}{2}^{+}}
=M_{\Sigma^{\ast }}^2+M_{\Omega^{}}^2-2M_{\Xi^{\ast }}^2
=M_{\Sigma_c^{\ast }}^2+M_{\Omega_c^{\ast }}^2-2M_{\Xi_c^{\ast }}^2
=M_{\Sigma_b^{\ast }}^2+M_{\Omega_b^{\ast }}^2-2M_{\Xi_b^{\ast }}^2
 \label{ns-3+} .
\end{equation}
   When $i=s$, $j=b$, $q=n,s,b$, Eqs. (\ref{baryon-equal}) can be expressed as
\begin{equation}
\delta_{sb,q}^{\frac{3}{2}^{+}}
=M_{\Xi^{\ast }}^2+M_{\Xi_{bb}^{\ast }}^2-2M_{\Xi_b^{\ast }}^2
=M_{\Omega }^2+M_{\Omega_{bb}^{\ast }}^2-2M_{\Omega_b^{\ast }}^2
=M_{\Omega_b^{\ast }}^2+M_{\Omega_{bbb}^{}}^2-2M_{\Omega_{bb}^{\ast }}^2 \label{sb-3+} .
\end{equation}

      Using Eqs. (\ref{regge1}) and (\ref{intercept2}),   one can obtain
\begin{equation} \label{combination}
\alpha_{iik}'M_{iik}^2+\alpha_{jjk}'M_{jjk}^2=2\alpha_{ijk}'M_{ijk}^2.
\end{equation}
      With Eqs. (\ref{slope1b}) and (\ref{combination}),  when the quark masses satisfy $m_j > m_i$, we can obtain \cite{Regge2008}
\begin{equation}
 \frac{\alpha_{jjk}'}{\alpha_{iik}'} = \frac{1}{2M_{jjk}^2}\times[(4M_{ijk}^2-M_{iik}^2-M_{jjk}^2)+\sqrt{(4M_{ijk}^2-M_{iik}^2-M_{jjk}^2)^2-4M_{iik}^2M_{jjk}^2} ]. \label{solution-b}
\end{equation}
    Therefore, for the $\frac{3}{2}^+$ multiplet,
\begin{equation}
\begin{aligned}  \label{s-bbs}
\frac{\alpha_{bbs}'}{\alpha_{nns}'} &= \frac{1}{2M_{\Omega_{bb}^{*-}}^2}\times[(4M_{\Xi_b^{*0}}^2-M_{\Sigma^{*+}}^2-M_{\Omega_{bb}^{*-}}^2)+\sqrt{(4M_{\Xi_b^{*0}}^2-M_{\Sigma^{*+}}^2-M_{\Omega_{bb}^{*-}}^2)^2-4M_{\Sigma^{*+}}^2 M_{\Omega_{bb}^{*-}}^2} ],  \\
\frac{\alpha_{bbs}'}{\alpha_{sss}'} &= \frac{1}{2M_{\Omega_{bb}^{*-}}^2}\times[(4M_{\Omega_{b}^{*-}}^2-M_{\Omega^-}^2-M_{\Omega_{bb}^{*-}}^2)+\sqrt{(4M_{\Omega_{b}^{*-}}^2-M_{\Omega^-}^2-M_{\Omega_{bb}^{*-}}^2)^2-4M_{\Omega^-}^2M_{\Omega_{bb}^{*-}}^2} ],  \\
\frac{\alpha_{sss}'}{\alpha_{nns}'} &= \frac{1}{2M_{\Omega^-}^2}\times[(4M_{\Xi^{*0}}^2-M_{\Sigma^{*+}}^2-M_{\Omega^-}^2)+\sqrt{(4M_{\Xi^{*0}}^2-M_{\Sigma^{*+}}^2-M_{\Omega^-}^2)^2-4M_{\Sigma^{*+}}^2 M_{\Omega^-}^2} ].
\end{aligned}
\end{equation}
    With the identical equation $\frac{\alpha_{bbs}'}{\alpha_{nns}'} \equiv \frac{\alpha_{bbs}'}{\alpha_{sss}'} \times \frac{\alpha_{sss}'}{\alpha_{nns}'}$,
  one can have
{\small
\begin{equation}
\begin{aligned}
 & \frac{1}{2M_{\Omega_{bb}^{*}}^2}\times[(4M_{\Xi_b^{*}}^2-M_{\Sigma^{*}}^2-M_{\Omega_{bb}^{*}}^2)+\sqrt{(4M_{\Xi_b^{*}}^2-M_{\Sigma^{*}}^2-M_{\Omega_{bb}^{*}}^2)^2-4M_{\Sigma^{*}}^2 M_{\Omega_{bb}^{*}}^2} ]   \\
=& \frac{1}{2M_{\Omega_{bb}^{*}}^2}\times[(4M_{\Omega_{b}^{*}}^2-M_{\Omega}^2-M_{\Omega_{bb}^{*}}^2)+\sqrt{(4M_{\Omega_{b}^{*}}^2-M_{\Omega}^2-M_{\Omega_{bb}^{*}}^2)^2-4M_{\Omega}^2M_{\Omega_{bb}^{*}}^2} ] \\
 & \times \frac{1}{2M_{\Omega}^2}\times[(4M_{\Xi^{*}}^2-M_{\Sigma^{*}}^2-M_{\Omega}^2)+\sqrt{(4M_{\Xi^{*}}^2-M_{\Sigma^{*}}^2-M_{\Omega}^2)^2-4M_{\Sigma^{*}}^2 M_{\Omega}^2} ].
\label{bbs}
\end{aligned}
\end{equation}
}

    In the following, we will give the mass expressions and mass values  for the ground state unobserved bottom baryons.   %
    Before this, we will apply the method first to the known $\Omega_{c}^{*}$  in order to see how well it works.
    From Eq. (\ref{ns-3+}), we can obtain the squared masse expression for $\Omega_{c}^{*}$  as a function of the squared masses of $\Sigma^{*}$, $\Xi^{*}$, $\Omega$,  $\Sigma_{c}^{*}$, and $\Xi_c^{*}$,
\begin{equation} \label{ssc}
M_{\Omega_c^{\ast }}^2 = M_{\Sigma^{\ast }}^2+M_{\Omega^{}}^2-2M_{\Xi^{\ast }}^2
 - M_{\Sigma_c^{\ast }}^2 + 2M_{\Xi_c^{\ast }}^2 .
\end{equation}
%
    From the latest RPP \cite{PDG2015},
 $M_{\Sigma^{*}} = 1385.0 \pm 2.3$ MeV,  $M_{\Xi^{*}} = 1533.4 \pm 1.7$ MeV,  $M_{\Omega^{}} = 1672.45 \pm 0.29$ MeV,
 $M_{\Sigma_{c}^{*}} = 2518.1 \pm 0.9$ MeV,  $M_{\Xi_c^{*}} = 2645.9 \pm 0.5$ MeV.
    Inserting   these mass values into the relation (\ref{ssc}),  one can have $M_{\Omega_c^{*}}$ = 2769.8 $\pm$ 2.6 MeV.
    In the latest RPP \cite{PDG2015}, $M_{\Omega_c^{*}}$ = 2765.9 $\pm$ 2.0 MeV.
  Our result for $M_{\Omega_c^{*}}$ is in reasonable agreement with the experimental value.
    In the following we will apply the method to bottom baryons.

    From Eq. (\ref{ns-3+}), we can obtain the squared mass of  $\Omega_{b}^{*}$ as a function of the squared masses of $\Sigma^{*}$, $\Xi^{*}$, $\Omega$,    $\Sigma_{b}^{*}$, and $\Xi_b^{*}$,
\begin{equation} \label{ssb}
M_{\Omega_b^{\ast }}^2 = M_{\Sigma^{\ast }}^2+M_{\Omega^{}}^2-2M_{\Xi^{\ast }}^2
 - M_{\Sigma_b^{\ast }}^2 + 2M_{\Xi_b^{\ast }}^2 .
\end{equation}
    From Eqs. (\ref{bbs}) and (\ref{ssb}), we can obtain the squared mass of $\Omega_{bb}^{*}$ as a function of the squared masses of $\Sigma^{*}$, $\Xi^{*}$, $\Omega$,  $\Sigma_{b}^{*}$, and $\Xi_b^{*}$,
%
    %
%
\begin{equation}
{ \scriptsize
\begin{aligned} \label{bbs-express}
 M^2_{\Omega_{bb}^{*}}= &
5 M_{\Xi_b^{*}}^2- 2 M_{\Xi^{*}}^2- M_{\Sigma_{b}^{*}}^2+ \frac{M_{\Omega}^2}{2}+ \frac{5 M_{\Sigma^{*}}^2}{2}-\frac{2 \left(M_{\Xi_b^{*}}^2-M_{\Sigma_{b}^{*}}^2\right) \left(M_{\Xi^{*}}^2-M_{\Sigma^{*}}^2\right)}{M_{\Omega}^2-2 M_{\Xi^{*}}^2+M_{\Sigma^{*}}^2}+
 \frac{\sqrt{M_{\Omega}^4 -2 M_{\Omega}^2 \left(4 M_{\Xi^{*}}^2+M_{\Sigma^{*}}^2\right)+ \left(M_{\Sigma^{*}}^2-4 M_{\Xi^{*}}^2\right){}^2}}{2 \left(M_{\Omega}^2-2 M_{\Xi^{*}}^2+M_{\Sigma^{*}}^2\right)}
\\
& \times \sqrt{\left(M_{\Omega}^2-2 M_{\Xi^{*}}^2+M_{\Sigma^{*}}^2\right) \left(M_{\Omega}^2-4 M_{\Xi_b^{*}}^2-2 M_{\Xi^{*}}^2-4 M_{\Sigma_{b}^{*}}^2+M_{\Sigma^{*}}^2\right)+4 \left(M_{\Xi_b^{*}}^2-M_{\Sigma_{b}^{*}}^2\right){}^2}
.
\end{aligned}
}
\end{equation}

    From  Eqs. (\ref{sb-3+}),  (\ref{ssb}), and (\ref{bbs-express}),
 we can obtain the mass expressions for   $\Xi_{bb}^{*}$  and $\Omega_{bbb}^{}$.
%
%
\begin{equation}
{ \scriptsize
\begin{aligned}
& M^2_{\Xi_{bb}^{*}}=
3M_{\Xi_b^{*}}^2 + M_{\Xi^{*}}^2 + M_{\Sigma_{b}^{*}}^2 + \frac{M_{\Sigma^{*}}^2}{2} -\frac{M_{\Omega}^2}{2}  -\frac{2 \left(M_{\Xi_b^{*}}^2-M_{\Sigma_{b}^{*}}^2\right) \left(M_{\Xi^{*}}^2-M_{\Sigma^{*}}^2\right)}{M_{\Omega}^2-2 M_{\Xi^{*}}^2+M_{\Sigma^{*}}^2}+
 \frac{\sqrt{M_{\Omega}^4 -2 M_{\Omega}^2 \left(4 M_{\Xi^{*}}^2+M_{\Sigma^{*}}^2\right)+ \left(M_{\Sigma^{*}}^2-4 M_{\Xi^{*}}^2\right){}^2}}{2 \left(M_{\Omega}^2-2 M_{\Xi^{*}}^2+M_{\Sigma^{*}}^2\right)}
\\
& \times \sqrt{\left(M_{\Omega}^2-2 M_{\Xi^{*}}^2+M_{\Sigma^{*}}^2\right) \left(M_{\Omega}^2-4 M_{\Xi_b^{*}}^2-2 M_{\Xi^{*}}^2-4 M_{\Sigma_{b}^{*}}^2+M_{\Sigma^{*}}^2\right)+4 \left(M_{\Xi_b^{*}}^2-M_{\Sigma_{b}^{*}}^2\right){}^2}
, \label{bbb-express}
\end{aligned}
}
\end{equation}
%
%
\begin{equation}
{ \scriptsize
\begin{aligned}
M^2_{\Omega_{bbb}^{}}= &
 9 M_{\Xi_b^{*}}^2 + \frac{9 M_{\Sigma^{*}}^2}{2} - \frac{M_{\Omega }^2}{2}-\frac{6 \left(M_{\Xi_b^{*}}^2-M_{\Sigma_{b}^{*}}^2\right) \left(M_{\Xi^{*}}^2-M_{\Sigma^{*}}^2\right)}{M_{\Omega}^2-2 M_{\Xi^{*}}^2+M_{\Sigma^{*}}^2}+
\frac{ \sqrt{-2 M_{\Omega}^2 \left(4 M_{\Xi^{*}}^2+M_{\Sigma^{*}}^2\right)+M_{\Omega}^4+\left(M_{\Sigma^{*}}^2-4 M_{\Xi^{*}}^2\right){}^2} }{2 \left(M_{\Omega}^2-2 M_{\Xi^{*}}^2+M_{\Sigma^{*}}^2\right)/3}
\\
& \times \sqrt{\left(M_{\Omega}^2-2 M_{\Xi^{*}}^2+M_{\Sigma^{*}}^2\right) \left(M_{\Omega}^2-4 M_{\Xi_b^{*}}^2-2 M_{\Xi^{*}}^2-4 M_{\Sigma_{b}^{*}}^2+M_{\Sigma^{*}}^2\right)+4 \left(M_{\Xi_b^{*}}^2-M_{\Sigma_{b}^{*}}^2\right){}^2}
. \label{bbb-express}
\end{aligned}
}
\end{equation}

   From the latest RPP \cite{PDG2015},
 $M_{\Sigma^{*}} = 1385.0 \pm 2.3$ MeV,  $M_{\Xi^{*}} = 1533.4 \pm 1.7$ MeV,  $M_{\Omega^{}} = 1672.45 \pm 0.29$ MeV,
 $M_{\Sigma_{b}^{*}} = 5833.6 \pm 2.4$ MeV,  $M_{\Xi_b^{*}} = 5952.1 \pm 3.3$ MeV.
    Inserting   the corresponding  mass values into the relation (\ref{ssb}),  one can have $M_{\Omega_b^{*}}$ = 6069.4 $\pm$ 6.9 MeV.
    Therefore, $M_{\Omega_b^{*}} - M_{\Omega_b^{}}$ = 21.4 $\pm$ 7.2 MeV,
  which is  close to the experimental data,  $M_{\Sigma_b^{*}} - M_{\Sigma_b^{}}$ = 21.2 $\pm$ 2.0 MeV and $M_{\Xi_b^{*-}} - M_{\Xi_b^{'-}}$ = 20.3 $\pm$ 0.1 MeV,  as it should be.
  The prediction, $M_{\Omega_b^{*}} - M_{\Omega_b^{}}$ = 21.4 $\pm$ 7.2 MeV, is in reasonable agreement with the value (24.0 $\pm$ 0.7 MeV) given  in Ref. \cite{Karliner:2008sv}.
  ~     Inserting the above mass values of $\Sigma^{*}$, $\Xi^{*}$, $\Omega$, $\Sigma_b^{*}$,   and $\Xi_b^{*}$ into the relation (\ref{bbs-express}), one can get $M_{\Omega_{bb}^{\ast}}$ = 10431 $\pm$ 40 MeV.  %
    From Eq. (\ref{bbb-express}) one has  $M_{\Omega_{bbb}^{}}$= 14788 $\pm$ 80 MeV.
   Similarly,  we can obtain the expression for $M_{\Xi_{bb}^{*}}$ and get its  value to be 10316.3 $\pm$ 37.8 MeV (10316 $\pm$ 38 MeV when truncated to the 1 MeV digit).
    Comparison of the masses of $\Omega_{b}^{*}$, $\Omega_{bb}^{*}$, $\Omega_{bbb}^{}$, and $\Xi_{bb}^{*}$ extracted in the present work and those given in other references is shown in Table 1.

{       

\begin{table}
 Table 1.   The masses of the ground state unobserved singly, doubly, and triply bottom baryons (in units of MeV).  Our results are labeled with ``Pre".
       The  isospin splittings, $M_{\Xi_{bb}^{-}} - M_{\Xi_{bb}^{0}}$ = $2.3 \pm 0.7 $ MeV, $M_{\Xi_{bb}^{*-}} - M_{\Xi_{bb}^{*0}}$ = $1.6 \pm 0.6$ MeV. \\
\begin{ruledtabular}
{   \footnotesize     
 \renewcommand\arraystretch{0.95}
\begin{tabular}{r| l|  *{4}{l} | l}     

                                      &  $\Omega_{b}^*$         &  $\Xi_{bb}^{}$                &  $\Omega_{bb}$              &  $\Xi_{bb}^{*}$               &  $\Omega_{bb}^*$                   &  $\Omega_{bbb}$           \\      \hline

Pre                             &  $6069.4 \pm 6.9$        &  10199$\pm$37                     & 10320$\pm$37               &  10316$\pm$38              & 10431$\pm$40                    & 14788$\pm$80                   \\      \hline    

\cite{Wang:2010vn}       		        &  6170$\pm$150         & 10170$\pm$140               &  10320$\pm$140              & 10220$\pm$150             &   10380$\pm$140              &   14830$\pm$100               \\      \hline

\cite{Ponce-1979}                         &  6051         &  10003                      &  10142                      & 10048                      &   10181                     &   14248                             \\      \hline

\cite{Martynenko:2007je}                  &  6102         &  10130                      &  10422                      &  10144                    &  10432                          &  14569                              \\      \hline


\cite{Zhang:2008pm}                  &  6000$\pm$160         &   9780$\pm$70               &  9850$\pm$70                &  10350$\pm$80                &    10280$\pm$50               & 13280$\pm$100                          \\      \hline

\cite{Roberts:2007ni}                     &  6102         &  10340                      &  10454                      &  10367                    &  10486                      &     14834                           \\      \hline

\cite{Kiselev-etal-2000-2002}             & ~~~~     &   10093                     &   10133                     &  10180                    & 10200                      &                                \\      \hline

\cite{Ebert:2005xj}                 &  6088         &   10202                     &   10359                     &  10237                    & 10389                      &                                \\      \hline

\cite{Bjorken:1985ei}                     &  6035$\pm$60         &                             &                             &  10250$\pm$120            &  10395$\pm$120              &   14760$\pm$180                             \\      \hline

\cite{Roncaglia:1995az}                   &  6090$\pm$50         &   10340$\pm$100             &   10370$\pm$100             &  10370$\pm$100            & 10400$\pm$100              &                                \\      \hline

\cite{He:2004px}                          &  ~~~~         &   10272                     &   10369                     &  10337                    & 10429                      &                                \\      \hline


\cite{Kiselev:2000jb}                     &  ~~~~         &   10000$\pm$80              &   10090$\pm$70              &                            &                             &                                \\      \hline

\cite{Jia-2006}                           &  ~~~~         &                             &                             &                            &                               &  14370$\pm$80                    \\      \hline
\cite{Hasenfratz:1980ka}                      &  ~~~~         &                             &                             &                            &                              &   14300                             \\      \hline





\cite{Brown:2014ena}        	        & 6085$\pm$47$\pm$20         & 10143$\pm$30$\pm$23       & 10273$\pm$27$\pm$20        & 10178$\pm$30$\pm$24    & 10308$\pm$27$\pm$21     & 14366$\pm$9$\pm$20                \\      \hline
\cite{Brown:2014pra}        	        & 6036$\pm$80         &   10185$\pm$53              &  10250$\pm$51               &    10191$\pm$56           &    10283$\pm$51             &  14370$\pm$10                             \\      \hline
\cite{Karliner:2014gca}     	        &  ~~~~         &   10162$\pm$12              &                             &    10184$\pm$12           &                             &                                \\      \hline
\cite{Ghalenovi:2014swa}    	        &  5986,6135           &   10334,10467           &  10397,10606             &    10431,10525           &    10495,10664                    &  15023,15175                          \\      \hline
\cite{Aliev:2012iv}         			    &  ~~~~         & 9960$\pm$90                 & 9970$\pm$90                  &    10300$\pm$200         &    10400$\pm$200            & 14300$\pm$200                 \\      \hline


\cite{Weng:2010rb}                   &  ~~~~         &   10080                     &  10100                      &                                &                             &                                \\      \hline

\cite{Bagan:1992za}                  &  ~~~~         &   9940$\pm$910              &                             &  10330$\pm$1090                &                             &                                \\      \hline

\cite{Patel:2008mv}                  &  6028-6132    &   9998-10137                &  10154-10269                &  10053-10206                   &  10228-10355                &  14444-14688                    \\      \hline

\cite{Ghalenovi:2011zz}              &  6096         &  10339,10344                &  10478                      &  10468,10473                   &  10607                     &  15118                       \\      \hline
\cite{Bernotas:2008bu}               &  6096         &  10062                      &  10208                      &  10101                         &  10244                      &   14276                       \\      \hline
\cite{Kiselev:2001fw}                &  ~~~~         &  10090                      &  10180                      &  10130                         &  10200                      &                                \\      \hline

\cite{Tang:2011fv}                   &  ~~~~         &  9800                       &  9890                       &  9840                          & 9930                       &                                \\      \hline
\cite{Giannuzzi:2009gh}              &  ~~~~         &   10185                     &  10271                      &  10216                         &  10289                      &                                \\      \hline

\cite{Valcarce:2008dr}               &  6079         &   10189                     &  10293                      &  10218                         & 10321                       &                              \\      \hline

\cite{LlanesEstrada:2011kc}          &  ~~~~         &                             &                             &                                &                             &  14200-15670              \\      \hline


\cite{Yoshida:2015tia}           &  6094         &   10314                     &   10447                    &    10339                      &  10467                     &                                \\      \hline

\cite{SilvestreBrac:1996bg}        &  ~~~~         & 10194                         & 10267                       &                                 &                             & 14398                          \\      \hline

\cite{Lewis:2008fu}              &6044$\pm$18$^{+20}_{-21}$   & 10127$\pm$13$^{+12}_{-26}$      & 10225$\pm$9$^{+12}_{-13}$   & 10151$\pm$14$^{+16}_{-25}$      & 10246$\pm$10$^{+18}_{-12}$    &                         \\      \hline

\cite{Zheng:2010zzc}                  &  ~~~~         &                             &                             & 9858                          & 10088                            & 14440                 \\      \hline

\cite{Albertus:2006ya}             &  ~~~~         & 10197$^{+10}_{-17}$           & 10260$^{+14}_{-34}$        & 10236$^{+9}_{-17}$               & 10297$^{+5}_{-28}$            &                         \\      \hline

\cite{AliKhan:1999yb}             &6069$\pm$34($^{-18}_{+30}$)($^{+35}_{-0}$)    & 10314$\pm$46$^{-10}_{+11}$     & 10365$\pm$40($^{-11}_{+12}$)($^{+16}_{-0}$)        & 10333$\pm$55$^{-7}_{+6}$               & 10383$\pm$39($^{-8}_{+8}$)($^{+12}_{-0}$)            &                         \\      \hline

\cite{Ebert:1996ec}             &  ~~~~         & 10230                            & 10320                         & 10280                         & 10360                     &                         \\      \hline

\cite{Tong:1999qs}             &  ~~~~         & 10300                            & 10340                         & 10340                         & 10380                     &                         \\      \hline

\cite{Narodetskii:2001bq}      &  ~~~~         & 10160                            & 10340                         &                           &                       &                         \\      \hline

\cite{Dhir:2013nka}            &  6142         & 10440                            & 10620                         & 10451                     &  10628                & 15129                   \\      \hline

\cite{Faessler:2006ft}         &  ~~~~         & 10100                            & 10180                         &                           &                       &                         \\      \hline

\cite{Garcilazo:2007eh}        & 6090         &&&&         &     \\      \hline

\cite{Karliner:2008sv}       & 6082.8$\pm$5.6         &&&&         &     \\      \hline

\cite{Jenkins:2007dm}       & 6058.9$\pm$8.1         &&&&         &     \\   

\end{tabular}
}
\end{ruledtabular}
\end{table}

}

       \subsection{  Masses of the ground unobserved   $\frac{1}{2}^+$    $\Omega_{bb}^{}$ and $\Xi_{bb}^{}$  baryons}


    Based on  Eqs.  (\ref{ns-3+})  and (\ref{sb-3+}), one can have
\begin{equation}
(M_{\Omega_{bb}^{*}}^2-M_{\Xi_{bb}^{*}}^2)+(M^2_{\Xi^{*}}-M^2_{\Sigma^{*}})=(M_{\Omega_b^{*}}^2-M_{\Sigma_b^{*}}^2) \label{Omega-bb-n3+} .
\end{equation}
   For the $\frac{1}{2}^+$ multiplet,  one can have a mass relation similar to  Eq. (\ref{Omega-bb-n3+}),
\begin{equation}
(M_{\Omega_{bb}^{}}^2-M_{\Xi_{bb}^{}}^2)+(M^2_{\Xi}-M^2_{\Sigma})=(M_{\Omega_b}^2-M_{\Sigma_b}^2) \label{Omega-bb-n1+} .
\end{equation}
   The linear forms in Eqs. (\ref{Omega-bb-n3+})  and (\ref{Omega-bb-n1+})  can satisfy
 the instanton model \cite{instanton-model} and the SU(8) symmetry \cite{SU8-Hendry-Verma}.
%

   For the $\frac{1}{2}^+$ multiplet, based on Eq. (\ref{baryon-equal}),
  $\delta_{nb,s}^{\frac{1}{2}^{+}} + \delta_{sb,n}^{\frac{1}{2}^{+}}$ can be expressed as
\begin{equation}
\delta_{nb,s}^{\frac{1}{2}^{+}} + \delta_{sb,n}^{\frac{1}{2}^{+}}
= M^2_{\Sigma}+M^2_{\Omega_{bb}} - 2\left(\frac{3M_{\Xi_b}^2+M_{\Xi_b'}^2}{4}\right)
+ M^2_{\Xi}+M^2_{\Xi_{bb}} - 2\left(\frac{3M_{\Xi_b}^2+M_{\Xi_b'}^2}{4}\right)
 \label{subeq:uc+dc-1+} .
\end{equation}
    In Ref. \cite{Regge2008},  it was pointed out that the values of $\delta_{ij}$ are only a little different between different multiplets for the same $i$ and $j$, although $\delta_{ij}$ are very sensitive to different quark flavors $i$ or $j$.
%
    As done in Refs. \cite{Regge2008, Wei:2015gsa}, assuming that $\delta_{ij}^{\frac{1}{2}^+} = \delta_{ij}^{\frac{3}{2}^+}$,
  we have $\delta_{nb}^{\frac{1}{2}^+} + \delta_{sb}^{\frac{1}{2}^+}$= $\delta_{nb}^{\frac{3}{2}^+} + \delta_{sb}^{\frac{3}{2}^+}$.
    The expression for $\delta_{sb}^{\frac{3}{2}^+}$ has been given in Eq. (\ref{sb-3+}).
     Based on Eq. (\ref{baryon-equal}), when $i=n$, $j=b$, $q=s$, $\delta_{nb,s}^{\frac{3}{2}^+}$ can be expressed as
\begin{equation}
\begin{aligned}
\delta_{nb,s}^{\frac{3}{2}^{+}}
=M_{\Sigma^{\ast }}^2+M_{\Omega_{bb}^{\ast }}^2-2M_{\Xi_b^{\ast }}^2
 \label{nb-3+}  .
\end{aligned}
\end{equation}
    With Eqs. (\ref{sb-3+}), (\ref{Omega-bb-n1+}), (\ref{subeq:uc+dc-1+}), and (\ref{nb-3+}),
  we can  obtain  the expressions for  $M_{\Omega_{bb}}$    and $M_{\Xi_{bb}}$.
    Then,  inserting the masses of  $\Sigma$, $\Xi$, $\Sigma_b$, $\Xi_b$, $\Xi_b'$, $\Omega_b$,   $\Sigma^*$, $\Omega$,   $\Xi_b^{*}$ from RPP \cite{PDG2015}
  and $\Omega_b^*$,  ${\Omega_{bb}^{*}}$ obtained in Sec. II,
  we can get the following values:  $M_{\Omega_{bb}}$= $10320 \pm 37 $ MeV, $M_{\Xi_{bb}^{}}$ = $10199 \pm 37$ MeV.
    Comparison of the masses of $\Xi_{bb}$ and $\Omega_{bb}^{}$  extracted in the present work and those given in other references is also shown in Table 1.

      \subsection{  Isospin splitting for the ground     $\Xi_{bb}^{*}$ and $\Xi_{bb}^{}$ baryons}

    Considering the difference between  $u$-quark and $d$-quark (which is the difference between the isospin multiplet),  
  for $\frac{3}{2}^+$ baryons,  Eq. (\ref{Omega-bb-n3+}) can be expressed as  
\begin{equation}
\begin{aligned}
(M_{\Omega_{bb}^{*-}}^2-M_{\Xi_{bb}^{*0}}^2)+(M^2_{\Xi^{*0}}-M^2_{\Sigma^{*+}})=(M_{\Omega_b^{*-}}^2-M_{\Sigma_b^{*+}}^2)  , \\
(M_{\Omega_{bb}^{*-}}^2-M_{\Xi_{bb}^{*-}}^2)+(M^2_{\Xi^{*-}}-M^2_{\Sigma^{*-}})=(M_{\Omega_b^{*-}}^2-M_{\Sigma_b^{*-}}^2) \label{Omega-bb-ud3+} .
\end{aligned}
\end{equation}
     Similarly, considering the isospin breaking effects,  for $\frac{1}{2}^+$ baryons,  Eq. (\ref{Omega-bb-n1+}) can be expressed as
\begin{equation}
\begin{aligned}
(M_{\Omega_{bb}^{-}}^2-M_{\Xi_{bb}^{0}}^2)+(M^2_{\Xi^0}-M^2_{\Sigma^+})=(M_{\Omega_b^-}^2-M_{\Sigma_b^{+}}^2) ,\\  
(M_{\Omega_{bb}^{-}}^2-M_{\Xi_{bb}^{-}}^2)+(M^2_{\Xi^-}-M^2_{\Sigma^-})=(M_{\Omega_b^-}^2-M_{\Sigma_b^-}^2) \label{Omega-bb-ud1+} .
\end{aligned}
\end{equation}
    From Eq. (\ref{Omega-bb-ud3+}), we  can  obtain the expression for the  isospin splitting $M_{\Xi_{bb}^{*-}} - M_{\Xi_{bb}^{*0}}$  and get its  value to be $1.6\pm 0.6$ MeV (where the uncertainties come from the errors of the input data).
    From Eq. (\ref{Omega-bb-ud1+}), we  can also obtain the expression for the  isospin splitting $M_{\Xi_{bb}^{-}} - M_{\Xi_{bb}^{0}}$  and get its  value to be $2.3 \pm 0.7$ MeV.
%
%
    %
%
%

   \section{Regge parameters and Masses of the  orbitally excited singly, doubly, and triply bottom baryons}   

    In Sec. II, all the masses  of  unobserved  ground state $\frac{1}{2}^+$ and $\frac{3}{2}^+$  singly, doubly, and triply bottom baryons were given.
    In this section, we will   calculate the Regge parameters and the masses of the orbitally excited baryons lying on the $\frac{1}{2}^+$ and $\frac{3}{2}^+$  bottom baryon trajectories.

   \subsection{Regge parameters and Masses of the  orbitally excited singly, doubly, and triply bottom baryons lying on the  $\frac{3}{2}^+$  trajectories}   %

%
        With the help of  $\alpha_{nns}' = 2/(M_{\Sigma(2030)}^{2}-M_{\Sigma^{*}}^{2})$,   
   from Eq. (\ref{s-bbs}), 
  one can have  the expression for $\alpha_{bbs}'$ and get its value ($\alpha_{bbs}'$ = $0.175 \pm 0.014$ GeV$^{-2}$).
    Similarly, using the masses of baryons presented in Eq. (\ref{s-bbs}), with the aid of Eq. (\ref{slope1b}),
  one can have the expressions for  $\alpha^\prime_{bnn}$, $\alpha^\prime_{bsn}$, $\alpha^\prime_{bss}$, $\alpha^\prime_{bbn}$,   and $\alpha^\prime_{bbb}$.
    Their values are given as follows:
%
 $\alpha^\prime_{bnn} = 0.295 \pm 0.022$ GeV$^{-2}$,
 $\alpha^\prime_{bsn} = 0.294 \pm 0.020$ GeV$^{-2}$,
 $\alpha^\prime_{bss} = 0.292 \pm 0.019$ GeV$^{-2}$,
 $\alpha^\prime_{bbn} = 0.176 \pm 0.015$ GeV$^{-2}$,
 $\alpha^\prime_{bbb} = 0.125 \pm 0.011$ GeV$^{-2}$.
    The results are shown in Table 2.

\begin{table}
    Table 2.    Regge slopes (in units of GeV$^{-2}$) and   Regge intercepts of the ground state $\frac{3}{2}^+$ singly, doubly, and triply bottom baryons. \\
\begin{ruledtabular}
{\small     
\begin{tabular}{c | *{6}{c}}
                    & $bnn$ ($\Sigma_b^*$)   & $bsn$ ($\Xi_{b}^*$)   & $bss$ ($\Omega_{b}^*$)           & $bbn$ ($\Xi_{bb}^*$)     & $bbs$ ($\Omega_{bb}^*$)   & $bbb$ ($\Omega_{bbb}$)  \\ \hline

$\alpha^\prime$     &0.295$\pm$0.022      & 0.294$\pm$0.020     &0.292$\pm$0.019                      &0.176$\pm$0.015      & 0.175$\pm$0.014     &0.125$\pm$0.011  \\ \hline

\emph{a}(0)         &-8.55$\pm$0.74     &-8.91$\pm$0.71     &-9.26$\pm$0.68                             &-17.22$\pm$1.44    &-17.58$\pm$1.41    &-25.89$\pm$2.14    \\

\end{tabular}
}
\end{ruledtabular}
\end{table}

%
%
       From Eq. (\ref{regge1}), one can have Regge intercepts,    
\begin{equation}
 a(0) = J - \alpha^\prime M^2 . \label{a0}
\end{equation}
    For example, with the mass for ${\Sigma_{b}^{\ast}}$ and  the value of $\alpha_{bnn}'$ obtained above, from Eq. (\ref{a0}),
  $a_{bnn}(0)$ can be given to be $-8.55 \pm 0.74$.
    Similarly, Regge intercepts $a(0)$ for other  $\frac{3}{2}^+$   trajectories can be given from Eq. (\ref{a0}).
    The results are also shown in Table 2.
    ~   From Eq. (\ref{regge1}), one has masses for orbitally excited states,
\begin{equation}
 M_J^2 = [J - a(0)]/\alpha' . \label{M-J}
\end{equation}
   Then, with the values for  $\alpha_{bnn}'$ and $a_{bnn}(0)$ obtained above, from Eq. (\ref{M-J}),
 the masses of the orbitally excited baryons ($L$=1,2,3,4 while $J^P=\frac{5}{2}^-,\frac{7}{2}^+,\frac{9}{2}^-,\frac{11}{2}^+$) lying on the $\Sigma_{b}^{\ast}$ trajectories
 can be expressed as functions of masses of the well-established light baryons and singly bottom baryons.
  The numerical results  are  shown in Table 3.
    Similarly, the mass expressions for all the orbitally excited singly and doubly $\frac{3}{2}^+$ bottom baryons can be extracted.
    The numerical results are  shown in Tables 3  and 4, respectively. (Here and below, masses of  orbitally excited baryons are truncated to the 1 MeV digit.)
        The wave function of a baryon  is antisymmetric.
    For the triply bottom baryon $\Omega_{bbb}^{}$, the flavor wave function  is symmetric, the color wave function  is antisymmetric.
    Therefore, its spin-orbital wave function should be symmetric.
   So, spin-symmetry ($S_{3q}$=$3\over{2}$) requires orbital-symmetry.
  Therefore, the odd-parity $\Omega_{bbb}^{}$ baryons ($L$=1,3) can only have the spin-antisymmetry ($S_{3q}$=$1\over{2}$).
    This will lead to the real masses lower than our calculated results for such baryons.
  The above $J^P=\frac{5}{2}^-, \frac{9}{2}^- $  will be  changed to $J^P=\frac{3}{2}^-, \frac{7}{2}^- $ for $L$=1, 3  $\Omega_{bbb}^{}$ baryons, respectively.
    The numerical results are  shown in Table 5.

\begin{table}
{  \footnotesize     
    Table 3. The masses of the singly bottom baryons lying on the $\frac{3}{2}^+$ trajectories (in units of MeV). Our results are labeled with ``Pre".
    The numbers in boldface are the experimental values taken as the input.\\
}

{  \ttiny     
 \renewcommand\arraystretch{1.8}

\begin{tabular}{r| *{5}{l}| *{5}{l}| *{5}{l}} \hline \hline

                                & & &${\Sigma_b^*}$ &&                                   && &${\Xi_b^*}$ &&                                  &\multicolumn{5}{c}{${\Omega_b^*}$}       \\    \cline{2-6} \cline{7-11} \cline{12-16}
 $J^P$                          &$\frac{3}{2}^+$  &$\frac{5}{2}^-$  &$\frac{7}{2}^+$  &$\frac{9}{2}^-$ &$\frac{11}{2}^+$           &$\frac{3}{2}^+$  &$\frac{5}{2}^-$  &$\frac{7}{2}^+$  &$\frac{9}{2}^-$ &$\frac{11}{2}^+$          &$\frac{3}{2}^+$  &$\frac{5}{2}^-$  &$\frac{7}{2}^+$  &$\frac{9}{2}^-$ &$\frac{11}{2}^+$      \\  \hline
Pre                       &\bf{5833.6$\pm$2.4}  &6117$\pm$20  &6388$\pm$38  &6648$\pm$55    &6898$\pm$71                       &\bf{5952.1$\pm$3.3} &6232$\pm$21  &6499$\pm$38     &6756$\pm$54  &7003$\pm$68                      &6069.4$\pm$6.9    &6345$\pm$22      &6609$\pm$38   &6863$\pm$52    &7108$\pm$66       \\  \hline

 \cite{PDG2015}              &5833.6$\pm$2.4       & &         & &                        &5952.1$\pm$3.3        & &         & &            &       &   &         & &           \\  \hline

\cite{Ebert:2011kk}             &5834   &6084   &6260         &6459   &6635                 &5963 &6226  &6414      &6610 &6782                     &6088 &6334  &6517      &6713  &6884    \\ \hline

\cite{Wang:2010vn}              &5850$\pm$200    &    &          & &                        &6020$\pm$170    &    &         & &                     & 6170$\pm$150  &    &         & &         \\ \hline

\cite{Ponce-1979}               &5769         &    &         & &                            &5912    &    &          & &                            & 6051  &    &         & &        \\ \hline

\cite{Roberts:2007ni}           &5858   &6172   &6333         & &                           &5980 &6201  &6395         & &                          &6102 &6311  &6497         & &       \\ \hline

\cite{Ebert:2005xj}       &5834 &6083   &6262         & &                             &5963   &6218   &6390         & &                       &6088   &6336   &6502        & &      \\ \hline

\cite{Roncaglia:1995az}         &5850$\pm$40   &   &         & &                            &5980$\pm$40   & &         & &                          &6090$\pm$50 & &        & &         \\  \hline

\cite{Yoshida:2015tia}     &5845  &6144    &          & &                              & &  &          & &                                     &6094 &6345  &         & &              \\ \hline

 \cite{Mao:2015gya}        &  &5980$\pm$180    &         & &                           & &6180$\pm$160  &         & &                          & &6430$\pm$130  &         & &          \\ \hline


\cite{Garcilazo:2007eh}    &5844   &   &         & &                                   &5967   &   &         & &                               &6090   &   &          & &              \\  \hline
\cite{Karliner:2008sv}     &5832.7$\pm$2   &   &         & &                                   &5959$\pm$4   &   &         & &                               &6082.8$\pm$5.6   &   &          & &           \\  \hline

\cite{Capstick:1986bm}          &5805 &6090   &6340        &6540 &                          &  &         & &             &                          & &  &         & &                      \\
 \hline \hline

\end{tabular}
}
\end{table}

\begin{table} [h]  
 { \footnotesize      
 Table 4.   The masses of the doubly  bottom baryons lying on the $\frac{3}{2}^+$  $\Xi_{bb}^{*}$ and $\Omega_{bb}^{*}$  trajectories  (in units of MeV).
  Our results are labeled with ``Pre".  
}
 { \scriptsize      
\begin{ruledtabular}
   \renewcommand\arraystretch{1.4}
\begin{tabular}{r  | *{5}{l}  | *{5}{l}}
                    &&&${\Xi_{bb}^*}$&&                                       
                                    &\multicolumn{5}{c}{${\Omega_{bb}^*}$}                             \\  \cline{2-6}   \cline{7-11}

 $J^P$                & $\frac{3}{2}^+$              & $\frac{5}{2}^-$       &$\frac{7}{2}^+$       & $\frac{9}{2}^-$       &$\frac{11}{2}^+$   
                      & $\frac{3}{2}^+$               & $\frac{5}{2}^-$               &$\frac{7}{2}^+$                & $\frac{9}{2}^-$        &$\frac{11}{2}^+$                   \\ \hline                        

Pre               & 10316$\pm$38           & 10588$\pm$59            & 10853$\pm$80          & 11112$\pm$99          & 11365$\pm$118     
                  & 10431$\pm$40               & 10700$\pm$60          & 10964$\pm$80                 &  11221$\pm$99          & 11472$\pm$117         \\ \hline


\cite{Roberts:2007ni}   &10367     &10731    &10608    &     &                      
                        &10486    &10766    &10732    &     &             \\ \hline

\cite{Kiselev-etal-2000-2002}     &10133    &10580    &10510   &     &                  
                                  &10257   &10670    &10627   &   &                        \\ \hline

\cite{Ebert:2005xj}       &10237       &10661      &   &      &            
                                 &10389      &10798      \\ \hline


\cite{Yoshida:2015tia}    &10339   &10759     &    &    &        
                            &10467   &10808     &    &    &         \\ \hline

\cite{Eakins:2012jk}              &10352       &10695       &11011   &    &         \\

\end{tabular}

\end{ruledtabular}
}
\end{table}

\begin{table} [h]  
{\footnotesize
Table 5.   The masses of the  triply bottom baryons lying on the  $\frac{3}{2}^+$  $\Omega_{bbb}^{}$ trajectory  (in units of MeV).     
  Our results are labeled with ``Pre". \\ 
\begin{ruledtabular}
   \renewcommand\arraystretch{1.0}
\begin{tabular}{r |  *{5}{l}}
                                               &\multicolumn{5}{c}{${\Omega_{bbb}}$}\\   \cline{2-6}
 $J^P$                              & $\frac{3}{2}^+$               & $\frac{3}{2}^-$               &$\frac{7}{2}^+$                & $\frac{7}{2}^-$           &$\frac{11}{2}^+$                \\ \hline        

Pre                                & 14788$\pm$80               & 15055$\pm$101                &  15318$\pm$123                & 15577$\pm$143             & 15831$\pm$163           \\ \hline


\cite{Wang:2010vn}        &14830$\pm$100       &14950$\pm$110       &       &       \\ \hline

\cite{Aliev:2012iv}             &14300$\pm$200       &14900$\pm$200      &   &      &                           \\ \hline

\cite{Roberts:2007ni}               &14834   &14976      &15101   &       &\\ \hline

\cite{Vijande:2014uma}          &14372      &14620     &14800                               \\ \hline
    \cite{Vijande:2015faa}          &14372      &14720     &14960                               \\ \hline




\cite{Meinel:2010pw}               & 14371$\pm$4$\pm$11$\pm$1     & 14714$\pm$29                         & 14969$\pm$40                   &         &                            \\

\end{tabular}

\end{ruledtabular}
}
\end{table}


   \subsection{Regge parameters and Masses of the  orbitally excited singly and doubly  baryons lying on the   $\frac{1}{2}^+$ trajectories}

    For the  $\frac{1}{2}^+$ trajectories, %
  according to the heavy quark symmetry  in the heavy quark limit,  Regge slopes of $\Sigma_b$, $\Xi_{b}'$, $\Omega_{b}^{}$, $\Xi_{bb}^{}$, and $\Omega_{bb}$  can be considered to be approximately equal to
                                those of  $\Sigma_b^{*}$, $\Xi_{b}^{*}$,  $\Omega_{b}^{*}$, $\Xi_{bb}^{*}$,   and $\Omega_{bb}^{*}$, respectively.
    According to  Refs. \cite{Burakovsky1997, Kobylinsky:1979zg}, $\alpha'_{\Lambda_b} \simeq \alpha'_{\Sigma_b}$, $\alpha'_{\Xi_{b}^{}} \simeq \alpha'_{\Xi_{b}'}$.
    We take these approximations in this work.
    Similar to the  $\frac{3}{2}^+$   trajectories,    from Eq. (\ref{a0}), one has  Regge intercepts $a(0)$ for $\frac{1}{2}^+$   trajectories.
    The numerical results are shown in Table 6.

\begin{table}
    Table 6.     Regge intercepts and Regge slopes (in units of GeV$^{-2}$) of the ground state $\frac{1}{2}^+$ singly and doubly bottom baryons.
\begin{ruledtabular}
{\footnotesize  
\begin{tabular}{c | *{7}{c}}

                & $bnn$ ($\Lambda_b$)    & $bnn$ ($\Sigma_b$)        & $bsn$ ($\Xi_b$)   & $bsn$ ($\Xi_b^\prime$)   & $bss$ ($\Omega_b$)        & $bbn$ ($\Xi_{bb}$)    & $bbs$ ($\Omega_{bb}$)     \\  \hline

$\alpha^\prime$     &0.295$\pm$0.022   &0.295$\pm$0.022              &0.294$\pm$0.020         & 0.294$\pm$0.020         &0.292$\pm$0.019                  &0.176$\pm$0.015   & 0.175$\pm$0.014     \\ \hline

\emph{a}(0)         &-8.82$\pm$0.68     &-9.48$\pm$0.73                 &-9.36$\pm$0.68         &-9.85$\pm$0.71         &-10.19$\pm$0.69                    &-17.80$\pm$1.41     &-18.17$\pm$1.39   \\
\end{tabular}
}
\end{ruledtabular}
\end{table}

    With the values for $a(0)$ and $\alpha^\prime$ listed in Table 6, from Eq. (\ref{M-J}),
   the masses of   the  orbitally excited baryons ($L$=1,2,3,4 while $J^P=$ $\frac{3}{2}^{-}$, $\frac{5}{2}^{+}$, $\frac{7}{2}^{-}$, $\frac{9}{2}^{+}$) lying on the  $\frac{1}{2}^+$  trajectories
  can be extracted.    
    The numerical results are  shown in Tables 7  and 8.


\begin{table}
    { \small
Table 7.  The masses of the singly bottom baryons lying on the $\frac{1}{2}^+$ trajectories (in units of MeV).
  Our results are labeled with ``Pre".
  The numbers in boldface are the experimental values taken as the input. \\
    }
\begin{ruledtabular}
{   \scriptsize  
 \renewcommand\arraystretch{1.4}
\begin{tabular}{r| *{5}{l}| *{5}{l} }
                            && &${\Lambda_b}$ &&                                                            &\multicolumn{5}{c}{${\Xi_b}$}             		             \\  \cline{2-6} \cline{7-11}
  $J^P$                     &$\frac{1}{2}^+$  &$\frac{3}{2}^-$  &$\frac{5}{2}^+$  &$\frac{7}{2}^-$ &$\frac{9}{2}^+$         &$\frac{1}{2}^+$  &$\frac{3}{2}^-$  &$\frac{5}{2}^+$  &$\frac{7}{2}^-$ &$\frac{9}{2}^+$   \\  \hline

Pre                    &\bf{5619.51$\pm$0.23}   &5913$\pm$21  &6193$\pm$40    &6461$\pm$58     &6718$\pm$74                    &\bf{5793.1$\pm$1.8}       &6080$\pm$19       &6354$\pm$37   &6616$\pm$53       &6869$\pm$68                          		      \\  \hline

 \cite{PDG2015}          &5619.51$\pm$0.23     &5919.73$\pm$0.32    &    &  &                           &5793.1$\pm$1.8    &    &                           		\\
                             \hline

\cite{Ebert:2011kk}         &5620   &5942  &6196           &6411  &6599                     &5803   &6130   &6373       &6581   &6762        		\\ \hline

\cite{Wang:2010vn}          &5650$\pm$200    &      &          &  &                          &5730$\pm$180    &    &                     		\\ \hline

\cite{Roberts:2007ni}       &5612       &5941       &6183         &  &                       &5806   &6093   &6300                       		\\ \hline


\cite{Ebert:2005xj}   &5622       &5947   &6197          &6405  &                     &5812   &6130   &6365       &6558                 		\\ \hline

\cite{Yoshida:2015tia}      &5618       &5939     &6212          &  &                        &   &     &                                 		\\ \hline

 \cite{Mao:2015gya}         &       &5880$\pm$110    &           &  &                        &       & 6070$\pm$130     &                		\\ \hline
\cite{Chen:2014nyo}         &5619   &5920   &6153          &6351  &6526                     &5801   &6106   &6349       &6559  &6747            		\\ \hline
\cite{Garcilazo:2007eh}     &5624   &5890     &          &  &                                &5825   &6076     &                         		\\      \hline
\cite{Karliner:2008sv}      & 5619.7$\pm$1.7   &5940$\pm$2      &      &      &             &5790-5800  &6115$\pm$4        &    &  &                 \\ \hline

\cite{Capstick:1986bm}      &5585   &5920  &6165         &6360  &6580                       &       &   &                               		\\

\end{tabular}
}
\end{ruledtabular}
\end{table}

\begin{table}
\vspace{-0.3 cm}
    { \footnotesize    Table 7.  Continued. \\   }
{  \ttiny  
 \renewcommand\arraystretch{1.8}
\begin{tabular}{r| *{5}{l}| *{5}{l}| *{5}{l}} \hline \hline

                                & & &${\Sigma_b}$ &&                                                             & & &${\Xi_b^\prime}$ &&                                                         & \multicolumn{5}{c}{${\Omega_b}$}                                     \\ \cline{2-6} \cline{7-11} \cline{12-16}
   $J^P$                        &$\frac{1}{2}^+$  &$\frac{3}{2}^-$  &$\frac{5}{2}^+$  &$\frac{7}{2}^-$ &$\frac{9}{2}^+$         &$\frac{1}{2}^+$  &$\frac{3}{2}^-$  &$\frac{5}{2}^+$  &$\frac{7}{2}^-$ &$\frac{9}{2}^+$         &$\frac{1}{2}^+$  &$\frac{3}{2}^-$  &$\frac{5}{2}^+$  &$\frac{7}{2}^-$ &$\frac{9}{2}^+$   \\  \hline

Pre                       &\bf{5813.4$\pm$2.8}     &6098$\pm$20   &6369$\pm$39  &6630$\pm$56  &6881$\pm$72                     & \bf{5935.0$\pm$0.5}  &6215$\pm$19   &6486$\pm$34  &6745$\pm$49  &6994$\pm$63                         &\bf{6048.0$\pm$1.9}   &6325$\pm$18    &6590$\pm$34    &6844$\pm$48    &7090$\pm$62                \\ \hline

 \cite{PDG2015}              &5813.4$\pm$2.8     &    &     &  &                       &5935.0$\pm$0.5     &  &     &  &                               &6048.0$\pm$1.9     &   &   &  &   \\
 \hline

\cite{Ebert:2011kk}             &5808       &6096   &6284     &6500  &6687               &5936   &6234   &6432   &6641  &6821                           &6064   &6340   &6529        & 6736  &6915  \\ \hline

\cite{Wang:2010vn}              &5800$\pm$190   &6000$\pm$180   &    &  &               &5960$\pm$170   &6140$\pm$140  &          &  &                   &6110$\pm$160   &6260$\pm$150    &         &  &  \\ \hline

\cite{Roberts:2007ni}           &5833       &6101   &6325          &  &                   &5970       &6190   &6393         &  &                           &6081   &6304 &6492    &  &    \\ \hline


\cite{Ebert:2005xj}       &5805       &6076   &6248          &  &                   &5937   &6212 &6377          &  &                                &6065   &6330 &6490     &  &    \\ \hline

\cite{Yoshida:2015tia}          &5823       &6132   &6397            &  &                 &       &       &              &  &                              &6076  &6336   &6561         &  &   \\ \hline

 \cite{Mao:2015gya}             &       & 5960$\pm$180   &         &  &                   &   & 6170$\pm$170   &          &  &                             & & 6430$\pm$130   &         &  &   \\ \hline
\cite{Garcilazo:2007eh}         &5789   &6039       &              &  &                   &5913   &6157   &              &  &                              &6037  &6278   &         &  &     \\  \hline
\cite{Karliner:2008sv}     & 5611.5$\pm$2   &5940$\pm$2      &      &      &             &5930$\pm$5  & & & &                          &6052.1$\pm$5.4      &    &  &           \\ \hline

\cite{Capstick:1986bm}          &5795       &6070   &6325       &6525  &                 &       &       &           &  &                                 &   & &         &  & \\
 \hline \hline

\end{tabular}
}

\end{table}

\begin{table} [h]  
    { \small
Table 8. The masses of the doubly bottom baryons lying on the  $\frac{1}{2}^+$ $\Xi_{bb}^{}$ and $\Omega_{bb}^{}$  trajectories (in units of MeV).   
    Our results are labelled with ``Pre".
    }
{  \scriptsize  
\begin{ruledtabular}
   \renewcommand\arraystretch{1.4}
\begin{tabular}{r | lllll    | lllll}
  && &${\Xi_{bb}}$ &&                             &\multicolumn{5}{c}{${\Omega_{bb}}$}               \\  \cline{2-6}  \cline{7-11}

 $J^P$                & $\frac{1}{2}^+$               & $\frac{3}{2}^-$               &$\frac{5}{2}^+$                & $\frac{7}{2}^-$                & $\frac{9}{2}^+$                        
                     & $\frac{1}{2}^+$               & $\frac{3}{2}^-$               &$\frac{5}{2}^+$                & $\frac{7}{2}^-$                & $\frac{9}{2}^+$   \\ \hline

Pre              & 10199$\pm$37              & 10474$\pm$59                 & 10742$\pm$80    & 11004$\pm$99               & 11259$\pm$118                                                 
                & 10320$\pm$37              &  10593$\pm$58                 & 10858$\pm$77                      & 11118$\pm$96                  & 11372$\pm$115        \\ \hline


\cite{Wang:2010vn}              &10170$\pm$140       &10390$\pm$150       &   &    &         
                               &10320$\pm$140        &10520$\pm$150    \\ \hline

\cite{Roberts:2007ni}       &10340   &10495    &10676   &    &                   
                             &10454   &10619   &10720   &  &     \\ \hline

\cite{Kiselev-etal-2000-2002}     &10093   &10343   &10497   &    &                    
                                 &10210        &10462    &    &    & \\ \hline

\cite{Ebert:2005xj}   &10202 &10408    &   &    &        
                           &10359  &10566   \\ \hline

\cite{Aliev:2012iv}         	    & 9960$\pm$90                 & 10430$\pm$150      &   &    &        
                             	    & 9970$\pm$90                 & 10570$\pm$150      &   &    &         \\ \hline


\cite{Yoshida:2015tia}    &10314       &10476    &10592   &    &        
                        &10447   &10608     &10729    &    &         \\ \hline

\cite{Eakins:2012jk}              &10322       &10692       &11002   &    &         \\

\end{tabular}
\end{ruledtabular}
}
\end{table}


\section{Discussion  and Summary}
\vspace{-3 mm}

    In the present work, we focused on  studying  masses of the unobserved singly, doubly, and triply bottom baryons.
    The results for the ground states  do not depend on  unobservable parameters (such as quark masses and Regge slopes) and distrustful resonances.
%
%
%
       Regge slopes and Regge intercepts of the   singly, doubly, and triply bottom baryon trajectories were extracted.
    After that, masses of the orbitally excited ($L$=1,2,3,4) singly, doubly, and triply bottom baryons were calculated.  
  ~
    In this work, all the input masses of baryons used in the calculations were taken from the latest ``Review of Particle Physics'' \cite{PDG2015}.   
    The uncertainties of the results only come from the errors of the input masses.
  Regge slopes and intercepts  used in this work were also calculated from  masses of light baryons and singly bottom baryons.
    No systematic error due to any small deviation from the Regge trajectories has been taken into account in this work.
%

    In Table 1,    comparison of the masses of ground state unobserved singly, doubly, and triply bottom baryons extracted in the present work and those given in other references is shown.
    In Table 2 and Table 6,  Regge parameters (slopes and intercepts) were given for the  $\frac{3}{2}^+$ and $\frac{1}{2}^+$ trajectories, respectively.
    In Tables 3, 4, and 5,   masses of   singly, doubly, and triply bottom baryons lying on the $\frac{3}{2}^+$ trajectories  are shown.
    In Table 7 and Table 8,    masses of   singly and doubly bottom baryons  lying on the   $\frac{1}{2}^+$ trajectories are shown, respectively.
    Our results are neither too big nor too small.

    Experimentally, there are two orbitally excited singly bottom baryons: $\Lambda_{b}(5912)$ and $\Lambda_{b}(5920)$ \cite{PDG2015},
 with $J^P$ = $\frac{1}{2}^-$, $\frac{3}{2}^-$ and $M_{\Lambda_{b}(5912)}$ = $5912.11 \pm 0.13 \pm 0.23$ MeV, $M_{\Lambda_{b}(5920)}$ = $5919.81 \pm 0.23$ MeV, respectively.
  $\Lambda_{b}(5912)$ and $\Lambda_{b}(5920)$ were first reported by LHCb Collaboration \cite{Aaij:2012da}.   
  $\Lambda_{b}(5920)$ was confirmed by CDF Collaboration \cite{Aaltonen:2013tta}.    
    The mass difference of these two states is small ($<8$ MeV).  This indicates that the spin-orbital couple has small impact on the  mass of the singly bottom baryon.
    In the present work, the mass of the $J^P$ = $\frac{3}{2}^-$ ($L$=1) singly bottom baryon is determined to be 5913 $\pm$ 21 MeV (shown in Table 1),
  which   consists with the experimental data.

    In Ref. \cite{Bjorken:1985ei}, Bjorken  pointed out $\frac{M_{\Omega_{bbb}}}{M_{\Upsilon}}$ =1.56$\pm$0.02 and gave the mass of $\Omega_{bbb}$ to be 14760$\pm$180 MeV,
 which consists with our result, $M_{\Omega_{bbb}} = 14788 \pm 80 $ MeV (shown in Table 1).
    In the present work, the central value of  mass splittings
 ($M_{\Omega_{bb}^{*}} - M_{\Xi_{bb}^{*}}$ = 10431 $-$ 10316 = 115 MeV,  $M_{\Omega_{bb}^{}} - M_{\Xi_{bb}^{}}$ = 10320 $-$ 10199 = 121 MeV)
  are reasonable (close to the usual constituent quark difference $m_s$ - $m_{u,d}$ $\approx$ 120 MeV).
    The central values of spin splittings  ($M_{\Omega_{bb}^{*}}-M_{\Omega_{bb}^{}}$ = 10431 $-$ 10320 = 111 MeV and $M_{\Xi_{bb}^{*}}-M_{\Xi_{bb}^{}}$ = 10316 $-$ 10199 = 117 MeV) are a little big.     
  However,   in Refs. \cite{Aliev:2012iv, Zhang:2008pm, Bagan:1992za}, the central values of these spin splittings  are even bigger than ours (more than 300 MeV).

    The isospin splitting for the $\frac{1}{2}^+$ doubly bottom baryons in the present work,  $M_{\Xi_{bb}^{-}} - M_{\Xi_{bb}^{0}}$ = $2.3 \pm 0.7 $ MeV,
  is smaller than  $5.3\pm 1.1$ MeV in Ref.  \cite{Hwang:2008dj} and $6.3\pm 1.7$ MeV in Ref. \cite{Brodsky:2011zs}.
        For  the $\frac{3}{2}^+$ doubly bottom baryons, we have also calculated the value of the  isospin splitting: $M_{\Xi_{bb}^{*-}} - M_{\Xi_{bb}^{*0}}$ = $1.6 \pm 0.6$ MeV.
    ~
    In  Ref. \cite{Brodsky:2011zs} strong and electromagnetic sources of isospin breaking are handled  separately.   
    Regge theory  appears to be a pure QCD emergent phenomenon.
  In this work, we do not consider the  electromagnetic corrections separately because the electromagnetic corrections cancel out  in Eqs. (\ref{sb-3+}) and (\ref{Omega-bb-ud1+}).
    These can be tested by experiments in the future.

      In Ref. \cite{Ebert:2011kk}, Ebert \emph{et al}. gave the values of Regge slopes for singly bottom baryons, which are a little bigger than the corresponding values in this work.
  For example, $\alpha'_{\Xi_{b}^{}}$ = 0.349 $\pm$ 0.019  GeV$^{-2}$ in Ref.  \cite{Ebert:2011kk}, while $\alpha'_{\Xi_{b}^{}}$ = 0.294 $\pm$ 0.020  GeV$^{-2}$ in this work.
    The mass of the orbitally excited state  decreases with the increase of the value of Regge slope.
    From  Tables 3 and 7, one can compare the masses given in  Ref.  \cite{Ebert:2011kk} and those given in the present work.
    The mass differences are small when $L$ = 1, 2.

   From Tables 1, 3-5, and 7-8, we can see that the masses of some ground state and orbitally excited  singly, doubly, and triply bottom baryons obtained here  are in good consistency with
 the existing experimental data and those given in many other different approaches. 
     However, our predictions for some highly orbitally excited baryons deviate considerably from some predictions in the literature.
     We expect that our predictions on the masses and Regge parameters in this manuscript can all be tested at LHCb in the near future.

    In this work, the squared mass relations were used rather than the linear mass relations taken in  Refs. \cite{{instanton-model},    {SU8-Hendry-Verma}}.
    When the mass relations include light baryons, bottom baryons, and doubly bottom baryons (such as Eqs. (\ref{sb-3+}) and (\ref{Omega-bb-ud1+})),
  the quadratic mass relations and the linear mass relations lead to significant different values.
    Besides   the addition for baryon spectra, searching for doubly and triply bottom baryons    
   can numerically check the quadratic mass relations and the linear mass relations.  
     The triply bottom baryons $\Omega_{bbb}$ are of considerable theoretical interest \cite{Jia-2006, Brambilla:2013vx},   
 since they are free of light quark contamination and may serve as a clean probe to the interplay between perturbative and nonperturbative QCD.
 %
    Therefore, more efforts should be given to study doubly and triply bottom baryons both theoretically and experimentally.

  \vspace{0.3 cm}
    The mass expressions, Regge parameters, and the  masse values in this work would be useful for the discovery of the unobserved singly, doubly, and triply bottom baryon states and
the $J^P$ assignment of these  baryon states when they are observed in the near future.

\begin{acknowledgments}
    One of us (Ke-Wei Wei) would like to thank Prof. Bing-Song Zou and Dr. Yao-Feng Zhang for support and discussions.
    This work was supported in part by National Natural Science Foundation of China (Project No. 11275025, No. U1204115, No. 11305003,  No. 11575023, No. 11605009, and  No. 11261130311)
  and  Key Project of Scientific and Technological Research of the Education Department of Henan Province (Project No. 12B140001 and No. 13A140014).
\end{acknowledgments}

\end{spacing}


\end{document}